\documentclass[conference, 10pt]{IEEEtran}
\usepackage{epsfig,rotating,setspace,latexsym,amsmath,epsf,amssymb,bm,theorem,amsfonts}
\usepackage{cite,subfigure,algorithmic}

\IEEEoverridecommandlockouts

\title{Information-Theoretic Analysis of an Energy Harvesting Communication System\thanks{This work
was supported by NSF Grants CCF 04-47613, CCF 05-14846, CNS 07-16311, CCF 07-29127 and CNS 09-64632.}}

\author{Omur Ozel \qquad Sennur Ulukus \\
\normalsize Department of Electrical and Computer Engineering\\
\normalsize University of Maryland, College Park, MD 20742 \\
\normalsize {\it omur@umd.edu} \qquad {\it ulukus@umd.edu}}

\newtheorem{theorem}{Theorem}

\begin{document}

\maketitle

\begin{abstract}
In energy harvesting communication systems, an exogenous
recharge process supplies energy for the data transmission and
arriving energy can be buffered in a battery before
consumption. Transmission is interrupted if there is not
sufficient energy. We address communication with such random
energy arrivals in an information-theoretic setting. Based on the classical additive white Gaussian noise (AWGN) channel model, we study the coding problem with random energy arrivals at the transmitter. We show that the capacity of the AWGN channel with stochastic energy arrivals is equal to
the capacity with an average power constraint equal to the
average recharge rate. We provide two different capacity achieving schemes: {\it save-and-transmit} and {\it best-effort-transmit}. Next, we consider the case where energy arrivals have time-varying average in a larger time scale. We derive the optimal offline power allocation for maximum average
throughput and provide an algorithm that finds the optimal power allocation.
\end{abstract}

\section{Introduction}

In this paper, we analyze point-to-point communication of
energy harvesting nodes from an information-theoretic
perspective. We focus on wireless networking applications where
nodes (e.g., sensors nodes) can harvest energy from nature
through various different sources, such as solar cells,
vibration absorption devices, water mills, thermoelectric
generators, microbial fuel cells, etc. In such systems, energy
that becomes available for data transmission can be modeled as
an exogenous recharge process. Therefore, unlike traditional
battery-powered systems, in these systems, energy is not a
deterministic quantity, but is a random process which varies
stochastically in time at a scale on the order of symbol
duration. The transmission can be interrupted due to lack of
energy in the battery. On the other hand, excess energy can be
buffered in the battery before consumption for transmission.
This model requires a major shift in terms of the power
constraint imposed on the channel input compared to those in
the existing literature.

To illustrate, in information-theoretic approaches, there are
two widely used input constraints on the channel inputs of
continuous-alphabet channels: average power constraint and
amplitude constraint. If the input is average power
constrained, then any codeword should be such that while each
symbol can take any real value, the average power of the entire
codeword should be no more than the power constraint. On the
other hand, if the input is amplitude constrained, then every
code symbol should be less than the constraint in amplitude. It
is clear that in an energy harvesting model, the constraint
imposed on the channel input is different than these constraints, in
that, while code symbols are instantaneously amplitude
constrained, energy can be saved in the battery for later use.
This amounts to an unprecedented input constraint on the
channel input. In this context, the main goal of this paper is
to investigate the effect of stochastic energy arrivals on the
communication in an information-theoretic framework. In
particular, we augment an energy buffer to the classical AWGN
system and study information-theoretically achievable rates.

First, we consider the setting where energy arrives at the transmitter as a stochastic process, which varies on the order of symbol duration. The problem is posed as design of a codebook
that complies with all of the input constraints. The input
constraint imposed by stochastic energy arrivals is a random
one and in this sense it generalizes classical deterministic
average or amplitude power constraints. The recharge process
together with past code symbols determine the allowable range
of inputs in each channel use. We start with showing that the capacity of the
AWGN channel with an average power constraint equal to the average
recharge rate is an upper bound for the capacity in the energy harvesting system. Then, we develop a scheme called {\it save-and-transmit} scheme that achieves this upper bound and hence the
capacity. The {\it save-and-transmit} scheme relies on sending zero code symbols in a portion of the total block length, which becomes negligible as the block length gets larger. Next, based on feasibility to send the code symbol in a channel use, we provide an alternative scheme called the {\it best-effort-transmit} scheme for achieving the capacity. Whenever the available energy is sufficient to send the code symbol, it is put to the channel, while a zero is put to the channel if there is not enough energy in the battery. We show that this scheme achieves rates arbitrarily close to the capacity.      

Secondly, we address a typical behavior in certain energy
harvesting sensors, such as solar-powered sensors, where the
recharge process is not ergodic or stationary. In this case, we
assume that the average recharging rate is not a constant in
time, but rather fosters time variation in a scale much larger
than the time scale that communication takes place.
Accordingly, we extend the formulation to the case in which
recharge process has a mean value that is varying in sufficiently long
time and we call each such sufficiently long time a slot. We
derive the optimal power control for maximum average throughput
in this case, and provide a geometric interpretation
for the resulting power allocation. We illustrate the advantage of the optimal power allocation in a numerical study.

\subsection{Relevant Literature}

To the best of our knowledge, this work is the first attempt to
analyze an energy harvesting communication system from an
information-theoretic perspective. There are many motivating
works in the networking literature. In \cite{Yates09TWC}, Lei
{\it et. al.} address replenishment in one hop transmission.
Formulating transmission strategy as a Markov decision process,
\cite{Yates09TWC} uses dynamic programming techniques for
optimization of transmission policy under replenishment. In \cite{tassiulas10TWC},
Gatzianas {\it et. al.} extend classical wireless network
scheduling results to a network with users having rechargeable batteries. Each battery is considered
as an energy queue, and data and energy queues are updated simultaneously in an interaction determined by
rate versus power relationship. Stability of data queues is studied using Lyapunov techniques. In \cite{sharma10TWC,sharma08Allerton}, in a similar energy harvesting setting, a dynamic power management policy is proposed and is shown to stabilize data queues. In each slot, energy spent is equal to the average recharge rate. Moreover, under a linear approximation, some delay-optimal schemes are proposed.

The capacity of scalar AWGN channel has been extensively
studied in the literature under different constraints on the
input signal. Average power (or the second moment) constraint
on the input yields the well-known result that the capacity
achieving input distribution is Gaussian with variance equal to
the power constraint. Smith \cite{Smith71, SmithThesis}
considers amplitude constraints in addition to average power
constraints and concludes that the capacity achieving input
distribution function is a step function with finite number of
increase points. Moreover, Shamai and Bar-David \cite{Shamai95}
extend Smith's result to amplitude constrained quadrature
Gaussian channel for which the optimal input distribution is
concentrated to finite number of uniform phase circles within
the amplitude constraint.

\section{AWGN Channel with Random Energy Arrivals} \label{model}

\begin{figure}
\begin{center}
\includegraphics[width=0.9\linewidth]{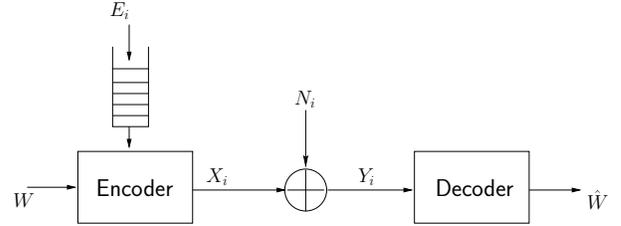}
\end{center}
\caption{AWGN channel with random energy arrivals.}
\label{sys_mod}
\end{figure}

We consider scalar AWGN channel characterized by the input $X$,
output $Y$, additive noise $N$ with unit normal distribution
$\mathcal{N}(0,1)$ and a battery (see Fig.~\ref{sys_mod}).
Input and output alphabets are taken as real numbers. 

Energy enters the system from a power source that supplies $E_i$ units
of energy in the $i$th channel use where $E_i\geq 0$.
$\{E_1,E_2,\ldots,E_n\}$ is the time sequence of supplied
energy in $n$ channel uses. $E_i$ is an i.i.d. sequence with
average value $P$, i.e., $E\left[E_i\right]=P$, for all $i$.

We assume that the energy stored and depleted from the battery are for only communication purposes (for example, the energy required for processing is out of the model). Existing energy in the battery can be retrieved without any loss and the battery capacity is large enough so that every quanta of incoming energy can be stored in the battery. This assumption is especially valid for the current state of the technology in which batteries have very large capacities compared to the rate of harvested energy flow \cite{sharma10TWC}: $E_{max}\gg P$. 

\begin{figure}
\includegraphics[width=1\linewidth]{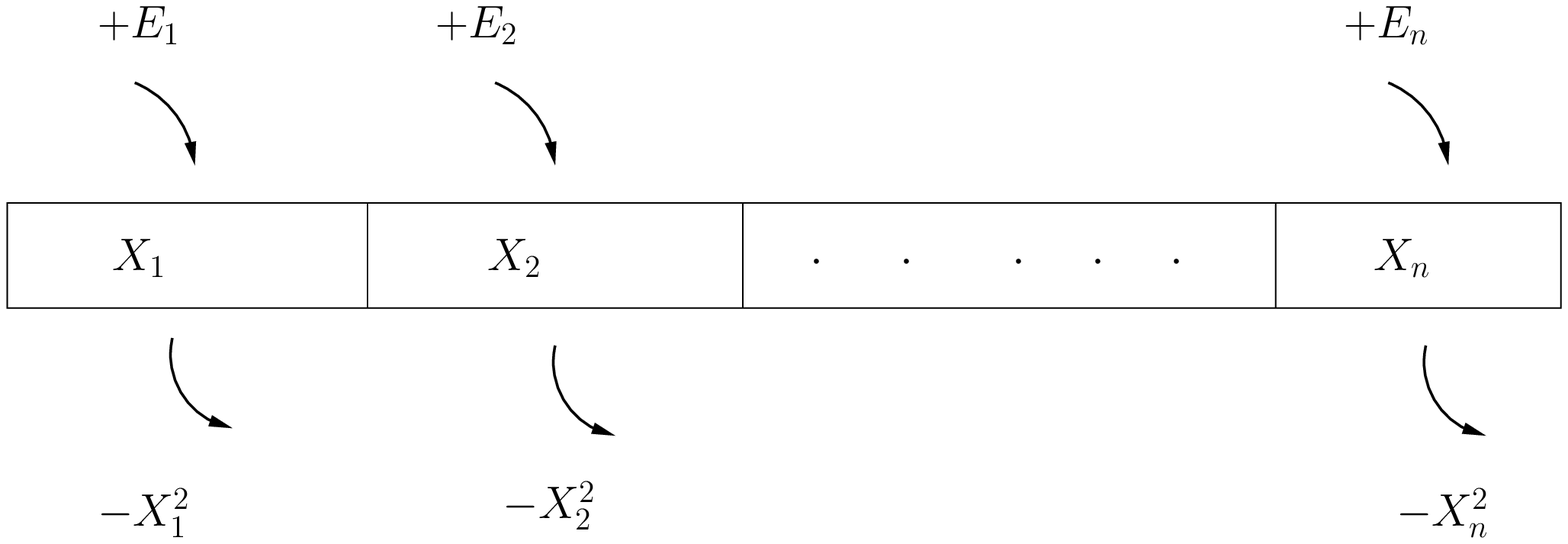}
\caption{For time $i$, $E_i$ denotes the energy arriving, and $X_i$ denotes
the channel input. Therefore, $X_i^2$ is the energy used at time $i$.}
\label{cw}
\end{figure}

The battery is initially empty and energy needed for communication
of a message is obtained from the arriving energy during
transmission of the corresponding codeword subject to
causality. In particular, $E_i$ units of energy is added to the
battery and $X_i^2$ units of energy is depleted from the
battery in the $i$th channel use. This brings us to the
following cumulative power constraints on the input based on
causality:
\begin{equation}
\label{constraints}
\sum_{i=1}^k X_i^2 \leq \sum_{i=1}^k E_i, \qquad k=1,\ldots,n
\end{equation}
This is illustrated in Fig.~\ref{cw}, where in each channel
use, $E_i$ amount of energy arrives and $X_i^2$ amount of
energy is used. Note that the constraints in
(\ref{constraints}) are upon the support set of the random
variables $X_i$. The first constraint restricts the support set
of $X_1$ to $[-\sqrt{E_1},\sqrt{E_1}]$. The second constraint
is $X_1^2 + X_2^2 \leq E_1 + E_2$. In general, letting $S_i$
denote $[\sum_{j=1}^{i-1} (E_j - X_j^2)]^+$, in channel use
$i$, the symbol $X_i$ is subject to the constraint $X_i^2 \leq
E_i + S_i$.

The input constraints in (\ref{constraints}) are challenging
because $E_i$ are random and these constraints introduce memory
(in time) in the channel inputs. Randomness in $E_i$ makes the
problem similar to fading channels in that state of the recharge
process (i.e., low or high $E_i$) affects instantaneous quality
of communication. Moreover, this time variation in recharge
process allows opportunistic control of energy as in fading
channels. However, recharge process can be collected in battery
unlike a fading state. In fact, we will see that, this nature
of energy arrivals makes it more advantageous to save energy in
the battery for future use when a peak occurs in the recharge
process, as opposed to opportunistically \textit{ride} the
peaks.

\section{The Capacity} \label{capacity}
We will invoke the general capacity formula of Verdu and Han
\cite{verduhan94}. For fixed $n$, let $f^n$ be the joint
density function of random variables $\{X_i\}_{i=1}^n$ and let
$\mathcal{F}^n$ be the set of $n$ variable joint density
functions that satisfy the constraints in (\ref{constraints}).
Since AWGN is an information-stable channel \cite{verduhan94},
the capacity of the channel in Fig.~\ref{sys_mod} with
constraints in (\ref{constraints}) is:
\begin{equation}
\label{cap2}
C = \lim_{n \rightarrow \infty} \frac{1}{n} \max_{f^n \in \mathcal{F}^n} I(X^n;Y^n)
\end{equation}
In general, capacity achieving input distribution is in the
form of product of marginal distributions (independent
distribution) \cite{verduhan94}. However, note that the power
constraints create dependence among the random variables. The
constraint on $X_{i+1}$ is dependent on given value of $X_j$,
$j\leq i$. Though independent processes achieve higher mutual
information than the ones with the same marginal distribution
but with correlation \cite{verduhan94}, the capacity that we
seek in this problem does not let the process be independent.
This problem falls in the family of problems of finding
capacity under dependence constraints on code symbols which is
by itself interesting and less studied.

An upper bound for $C$ is the corresponding AWGN capacity with
average power $P$, as $\frac{1}{n}\sum_{i=1}^n X_i^2 \leq
\frac{1}{n}\sum_{i=1}^n E_i$ and by the i.i.d. nature of $E_i$,
invoking strong law of large numbers \cite{Cover06},
$\frac{1}{n}\sum_{i=1}^n E_i \rightarrow P$ with probability 1.
Therefore, each codeword satisfying constraints in
(\ref{constraints}) automatically satisfies
$\lim_{n \rightarrow \infty}\frac{1}{n}\sum_{i=1}^n X_i^2 \leq P$ with probability 1.
However, if a codeword satisfies average power constraint, it does not necessarily satisfy the constraints in $(\ref{constraints})$. Hence, we get the
following bound:
\begin{equation}
\label{bound}
C \leq \frac{1}{2}\log\left( 1 + P\right)
\end{equation}
Next, we state and prove that the above upper bound can be
achieved.
\begin{theorem}
\label{cap} The capacity of the AWGN channel under i.i.d.
random energy arrivals $E_i$, where $E[E_i]=P$, is independent
of the realizations of $E_i$ and equal to the channel capacity
under average power constraint $P$:
\begin{equation}
C = \frac{1}{2} \log\left(1+P\right)
\end{equation}
\end{theorem}
To prove the theorem, we need an achievable scheme. Achieving
the capacity requires design of an $(n,2^{nR_n},\epsilon_n)$
codebook with encoding function $\{f^n_k(.)\}_{k=1}^n$ and
decoding function $g_n(.)$ where $n$ is the code length,
$2^{nR_n}$ is the code size and $\epsilon_n$ is the probability
of error such that $\epsilon_n \rightarrow 0$ and $R_n
\rightarrow C$.

There are two separate causes of error. The first one is that
any codeword does not satisfy the input constraints. In this
case, the encoder does not send that codeword but sends an all
zero codeword\footnote{Note that zero is costless (requires no
energy) and therefore all zero codeword satisfies the input
constraints for all realizations of the energy arrival
process.}. Hence, all zero codeword is assumed to be in the
codebook and it is decoded to message $0$ at the receiver. If
the decoder maps received signal to message $0$, it is accepted
as an error. Second cause of error is the actual decoding error
at the receiver. If the received signal is decoded to a message
that is different from the message sent, then an error occurs.
Accordingly, the error event is defined as union of these two
events.

While designing the codebook and the encoding/decoding rule, a
first approach can be to optimize the codebook design subject
to the input constraints. Therefore, the occurrence of the
first type of error (i.e., insufficient energy) is eliminated
from the beginning. However, we will show that, it is not
necessary to solve this difficult optimization problem.
Instead, the asymptotic behavior of the constraints allows us
to design codes that do not obey the input constraints for a
particular $n$ with a small probability but the requirement is
that small probability goes to zero asymptotically. For the
error probability to go to zero, we need to average out the
error due to randomness introduced by the channel and due to
the randomness in the energy arrivals. To do this, we propose a
scheme that implements a {\it save-and-transmit} principle in
that error due to randomness in energy is averaged out first
and then the channel coding is performed.

In addition, it is possible to maintain error-free communication even if some code symbols cannot be put to the channel correctly. Therefore, we also investigate achievable rates using a {\it best-effort-transmit} scheme, which can interestingly approach $\epsilon$ neighborhood of the capacity for any $\epsilon>0$. 

\section{The Save-and-Transmit Scheme}

We now present an achievable scheme. Let $h(n) \in o(n)$ such
that $\lim_{n \rightarrow \infty} h(n) = \infty$. Here, $o(n)$
denotes the class of functions $y(n)$ such that $\lim_{n
\rightarrow \infty} \frac{y(n)}{n} = 0$. Consider the
sequence of codes with code length $n$ and rate $R_n$ such that
the first $h(n)$ symbols of each codeword is zero and the
remaining $n - h(n)$ codewords are chosen as independent random
variables from the (capacity achieving) Gaussian distribution
with variance $P$ where $P$ is the average recharge rate. That
is, for $k=1,2,...,h(n)$, we have, the encoding function, $f_k^n(m) = 0$ for
all $m \in \{1,2,...,2^{nR}\}$.
For $k = h(n)+1,...,n$, $f_k^n(m)$ comes from
realizations of a Gaussian distributed random variable of
variance $P$ for all $m \in \{1,2,...,2^{nR}\}$.

\begin{figure}
\begin{center}
\includegraphics[width=1.0\linewidth]{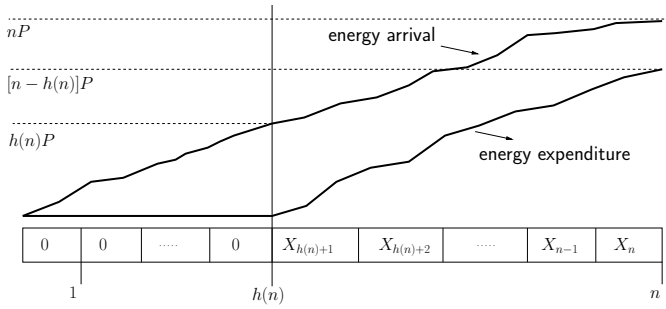}
\caption{The codewords in the save-and-transmit scheme. First $h(n)$ code
symbols are identically zero for all codewords. Remaining $n-h(n)$ code symbols
are selected as i.i.d. Gaussian distributed. Evolution of a sample energy
arrival and energy expenditure for a particular codeword is illustrated.
Collected energy in the first $h(n)$ channel uses makes it asymptotically
impossible to deem any codeword infeasible.}
\label{sys_mod3}
\end{center}
\end{figure}

As $n$ grows large, by strong law of large numbers, at time
index $h(n)$, about $h(n)P$ amount of energy is collected in
the battery with very high probability. Since no energy is
consumed up to this time, the codebook is feasible until time
$h(n)$. After time $h(n)$, the probability that a code symbol
requires more energy than the energy available in the battery
goes to zero and energy consumed is accounted for the energy
arrived. In the next $h(n)$ channel uses, $h(n)P$ units of
energy is collected in the battery and another $h(n)P$ is
consumed for data transmission. We repeat this procedure for a
total of $n/h(n)$ times, where in each interval of $h(n)$
symbols, we collect about $h(n) P$ units of energy and use
about $h(n) P$ units of energy collected in the previous block
of $h(n)$ symbols. The first collected amount guarantees that
the codewords are always feasible (see Fig.~\ref{sys_mod3}).
More specifically, as the length of the codewords gets large,
by strong law of large numbers, we have $\sum_{i=h(n)}^u X_i^2
\leq \sum_{i=1}^u E_i$ almost surely for all $u > h(n)$. For a precise proof of these arguments, we need to make use of Marcinkiewicz-Zygmund type strong laws for sums of i.i.d. random variables. Please see \cite{report10} for a complete proof.

The achievable rate for this scheme is
\begin{align}
\frac{1}{n}I(X^n;Y^n) & = \frac{1}{n}
\sum_{j=h(n)}^n I(X_j;Y_j) = \frac{n-h(n)}{2n}
\log\left(1 + P\right) \nonumber\\
& \rightarrow  \frac{1}{2} \log(1+P)
\end{align}
In other words, communication can start with $h(n)$ amount of
delay and the capacity of a corresponding average power constrained
AWGN channel can be achieved. Since, this scheme is asymptotically
feasible, this proves the achievability part and completes the
proof of Theorem~\ref{cap}.

It should be emphasized that the above achievable scheme is obtained without using causal or noncausal information of energy arrivals. For any realization of $E_i$, we introduce a reasonable delay $h(n)$ to save energy and afterwards transmit with average power $P$. This scheme obeys the input constraint with probability 1 and achieves the upper bound in (\ref{bound}). The ability to collect energy in the battery allows designer to apply a save-and-transmit strategy so that the uncertainty in the energy arrivals is eliminated first and then the uncertainty in the channel is dealt with by means of appropriate channel coding.

\section{The Best-Effort-Transmit Scheme}

Let $X^n = (X_1, X_2, \ldots,X_n)$ be a codeword of length $n$ where $X_i$ is the code symbol to be sent in channel use $i$ and the codebook be $\mathcal{C}^n$. The codebook that two parties agree upon is determined by generating independent Gaussian distributed random variables with mean zero and variance $P-\epsilon$ for each code symbol.  
Let $S(i)$ be the battery energy just before the $i$th channel use starts. In the best-effort-transmit scheme, the code symbol $X_i$ can be put to the channel if $S(i)\geq X_i^2$. Otherwise, transmitter puts a code symbol $0$ to the channel as battery does not have sufficient energy to send symbol $X_i$. Hence, the input to the channel is $X_i \textbf{1}(S(i) \geq X_i^2)$ and therefore there is a possible mismatch. Yet, we will show in the following that the mismatch does not affect the correct decoding of the message at the receiver.

The battery energy is updated as follows \begin{align} \label{updates} S(i+1)=S(i)+E_i-X_i^2 \textbf{1}(S(i)\geq X_i^2) \end{align}  
The energy updates in (\ref{updates}) are analogous to the queue updates in classical slotted systems \cite{tassiulas10TWC} where the notion of a slot is replaced with channel use. Therefore, battery acts like an energy queue. Unlike in the classical systems with packet queues, our aim is to keep the energy queue unstable so that the availability of the energy to send a code symbol is asymptotically guaranteed.  

The key to the updates in (\ref{updates}) is to put $0$ symbol on the channel as it guarantees that infeasible code symbols are observed only finitely many times, which is stated in the following theorem, and detailed proof is again skipped for brevity and can be found in \cite{report10}. 
\begin{theorem}
\label{fmthm} 
Let $X^n$ be a codeword such that $\{X_i\}$ is an i.i.d. real random sequence with mean zero and variance $P - \epsilon$ for some $\epsilon>0$ sufficiently small. Let $E_i$ be the energy arrivals and $S(i)$ be the battery energy which is updated as in (\ref{updates}). Then, only finitely many code symbols are infeasible. 
\end{theorem}

It is well known that we can pack about $2^{\frac{n}{2}\log\left(1 + (P-\epsilon)\right)}$ codewords with block length $n$, each having average power less than or equal to $P-\epsilon$, in the $\textbf{R}^n$ space such that, whenever one of them is sent by the transmitter, a joint typicality decoder \cite{Cover06} which checks whether a codeword is jointly typical with the received signal vector, can reconstruct the codeword at the receiver with probability of error approaching zero. In the {\it best-effort-transmit} scheme, the received signal vector will still be jointly typical with the sent codeword, since the number of infeasible code symbols is only finitely many due to Theorem \ref{fmthm}. Hence, the same decoder can recover the codeword if the channel input is $X_i \textbf{1}(S(i)\geq X_i^2)$ in each channel use. Therefore, the code rate $R=\frac{1}{2}\log\left(1 + (P-\epsilon)\right)$ is achievable, and hence by continuity of $\log(.)$, $R<\frac{1}{2}\log\left(1 + P\right)$ can be achieved.

\section{Optimal Power Control in a Large Time Scale}

We have seen that classical AWGN capacity with average power
constraint can be achieved with $o(n)$ delay for fueling the
battery with energy if the recharge process is i.i.d. However, the recharge process can deviate
from its i.i.d. characteristic in a large time scale. In
particular, the mean value of the recharge process may vary
after a long time. In the classical example of sensor nodes
fueled with solar power, mean recharge rate changes depending
on the time of the day. As an example, the mean recharge rate
may vary in one-hour slots and the sensor may be on for twelve
hours a day, in which case, a careful management of energy
expenditure in each slot will be required to optimize the
average performance during the day.

\begin{figure}
\includegraphics[width=1.0\linewidth]{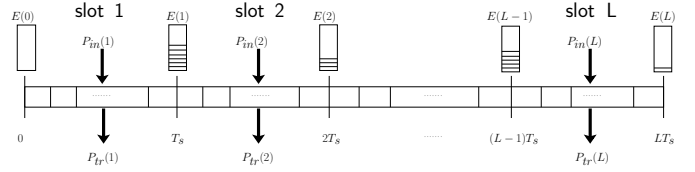}
\caption{$L$ large time slots. In each slot, sufficiently large time passes
to achieve AWGN capacity with average power constrained to allocated power
in that slot.}
\label{sys_mod4}
\end{figure}

We generalize the system model for $L$ large time slots (see
Fig.~\ref{sys_mod4}). We assume that the duration of each slot
is $T_s$ (large enough). For each slot $i=1,2,...,L$, average
recharge rate is $P_{in}(i)$ and $P_{tr}(i)$ units of power is
allocated for data transmission. In slot $i$, $P_{in}(i) T_s$
units of energy enters the battery and $P_{tr}(i) T_s$ units of
energy is spent for communication. If $P_{in}(i)
> P_{tr}(i)$, $(P_{in}(i)-P_{tr}(i))T_s$ units of energy is saved, or
otherwise $(P_{tr}(i)-P_{in}(i))T_s$ units of energy is
depleted from the battery. Assuming zero initial energy in the
battery and large enough battery capacity, every unit of incoming
energy is saved in the battery. The causality of energy
arrivals requires
\begin{equation}
\sum_{i=1}^\ell P_{tr}(i) \leq \sum_{i=1}^\ell P_{in}(i), \quad \ell=1,\ldots,L
\end{equation}

Around $\frac{1}{2}\log\left(1 + P_{tr}(i)\right) T_s$ bits of
data are sent in slot $i$. Before the communication starts,
suppose the designer knows the mean recharge rates $P_{in}(i)$
for all $i$, calculates $P_{tr}(i)$ and adjusts the average
power of codewords in slot $i$ to $P_{tr}(i)$ during
transmission\footnote{Changing the average
power of codewords requires using different codebooks in each
slot. However, scaling a common codebook by slot power
$P_{tr}(i)$ works as well. This can also be interpreted as a codebook with dynamic power allocation \cite{CS99} in slow time variation.}. We allocate transmit power to each slot subject to causality constraint so that
average throughput in $L$ slots is optimized:
\begin{eqnarray}
\mbox{max} & & \frac{1}{L}\sum_{i=1}^L \frac{1}{2}\log\left(1 + P_{tr}(i)\right) \nonumber \\
\mbox{s.t.} & & \sum_{i=1}^\ell P_{tr}(i) \leq \sum_{i=1}^\ell P_{in}(i), \quad \ell=1,\ldots, L
\end{eqnarray}
We will denote the solution of the above optimization problem as
$\textbf{P}_{tr}^*=[P_{tr}^*(1),P_{tr}^*(2),...,P_{tr}^*(L)]$.
Note that the slot duration $T_s$ does not appear in the
optimization problem. The rest of this section is devoted to
characterizing $\textbf{P}_{tr}^*$. After developing necessary
concepts, we present an algorithm to find $\textbf{P}_{tr}^*$
and illustrate it for several cases.

\subsection{Solution of the Optimization Problem}

In order to understand how the optimal solution may look like,
we will first pose the simplest version of the problem. Suppose
there is $cLT_S$ amount of energy available in the battery
where $c>0$ is a constant and the recharge process is zero.
Then, the optimal power allocation strategy $\textbf{P}_{tr}^*$
is the one that maximizes $\frac{1}{L} \sum_{i=1}^L
\frac{1}{2}\log\left(1 + P_{tr}(i)\right)$ subject to
$\sum_{i=1}^L P_{tr}(i) \leq cL$. By Jensen's inequality
\cite{Cover06}, optimal strategy is $P_{tr}^*(i)=c$. That is,
if $P_{tr}(i) \neq P_{tr}(j)$ for $i \neq j$, then we can
always achieve higher average throughput by setting $P'_{tr}(i)
= P'_{tr}(j) = \frac{P_{tr}(i)+P_{tr}(j)}{2}$. Returning to the
original problem, above observation translates to a general
optimality criterion for the problem, stated in the following
theorem. Please see \cite{report10} for a proof of this theorem.

\begin{theorem}
Let $\textbf{P}_{tr}=[P_{tr}(1),...,P_{tr}(L)]$ be any power
allocation such that $\sum_{i=1}^\ell P_{tr}(i) \leq
\sum_{i=1}^\ell P_{in}(i)$, for $\ell=1,\ldots,L$. Assume
$\textbf{P}'_{tr} \neq \textbf{P}_{tr}$ be another power
allocation such that $\sum_{i=1}^\ell P'_{tr}(i) \leq
\sum_{i=1}^\ell P_{in}(i)$, $\ell=1,\ldots,L$ and for some $e,s
\in \{1,2,..,L\}$ with $e<s$
\begin{equation*}
P'_{tr}(i)=\left\{\begin{array}{ll}
c, & \mbox{$i\in \{e,e+1,...,s\}$}\\
P_{tr}(i), & \mbox{$i\notin \{e,e+1,...,s\}$}.
\end{array}
\right.
\end{equation*}
Then, $\sum_{i=1}^L \frac{1}{2} \log\left(1 + P'_{tr}(i)\right)
> \sum_{i=1}^L \frac{1}{2} \log\left(1 + P_{tr}(i)\right)$.
\end{theorem}
This theorem proposes a method to optimize the power
allocation. It should be performed such that variation in power
from slot to slot is avoided as much as possible and the
energy should be consumed in as smooth a way as possible. We
will now make this more precise by describing the method of
finding the solution $\textbf{P}_{tr}^*$. 

It is inferred that $\textbf{P}_{tr}^*$ must take a constant value (not necessarily the same value) in each slot. In addition, we can partition the whole interval into disjoint intervals over which $P_{tr}^*(i)$ is constant  as follows:

\begin{equation}
P^*_{tr}(i)=\frac{\text{energy} \ \text{recharged}
\ \text{in} \ [n_{k}T_s, n_{k+1}T_s]}{(n_{k+1} - n_k)T_s}
\end{equation}
for all $i$ in $\{n_{k} + 1, ... , n_{k+1}\}$ for some subset
$\{n_k\}$ of $\{1,2,..,L\}$ with $n_{k+1} > n_k$.
Thus, the problem can also be cast as finding the right
subset $\{n_k^*\}$ that optimizes the average throughput. Note that $0$ and $L$ are always in this
subset. Hence, $n_1^* = 0$ and the element with last order is $L$. Define the cumulative energy
arrival rate as $e(i)=\sum_{j=1}^i P_{in}(j)$ for all $i \in \{1:L\}$
and reset $e(0)=0$. In each point, we have to use the
knowledge of the amount of energy that is going to arrive in
the remaining time interval and determine feasible constant
power strategies and find the optimal one among them. We have
$n_1^*=0$ and next $n_i^*$ are determined for optimal strategy as
follows \cite{report10}: \begin{equation} n_i^*=\arg\min_{k \in \{n_{i-1}^*+1:L\}}
\ \frac{e(k)-e(n_{i-1}^*)}{k-n_{i-1}^*}
\end{equation}
Then, the optimal power allocation is as follows:
\begin{equation}
P^*_{tr}(i)=\frac{e(n_i^*) - e(n_{i-1}^*)}{n_i^* - n_{i-1}^*}, \ \qquad
\text{for} \ \text{all} \ i \in \{n_{i-1}^*:n_i^*\}
\end{equation}

\begin{figure}[t]
\centering
\subfigure{
\includegraphics[width=0.46\linewidth]{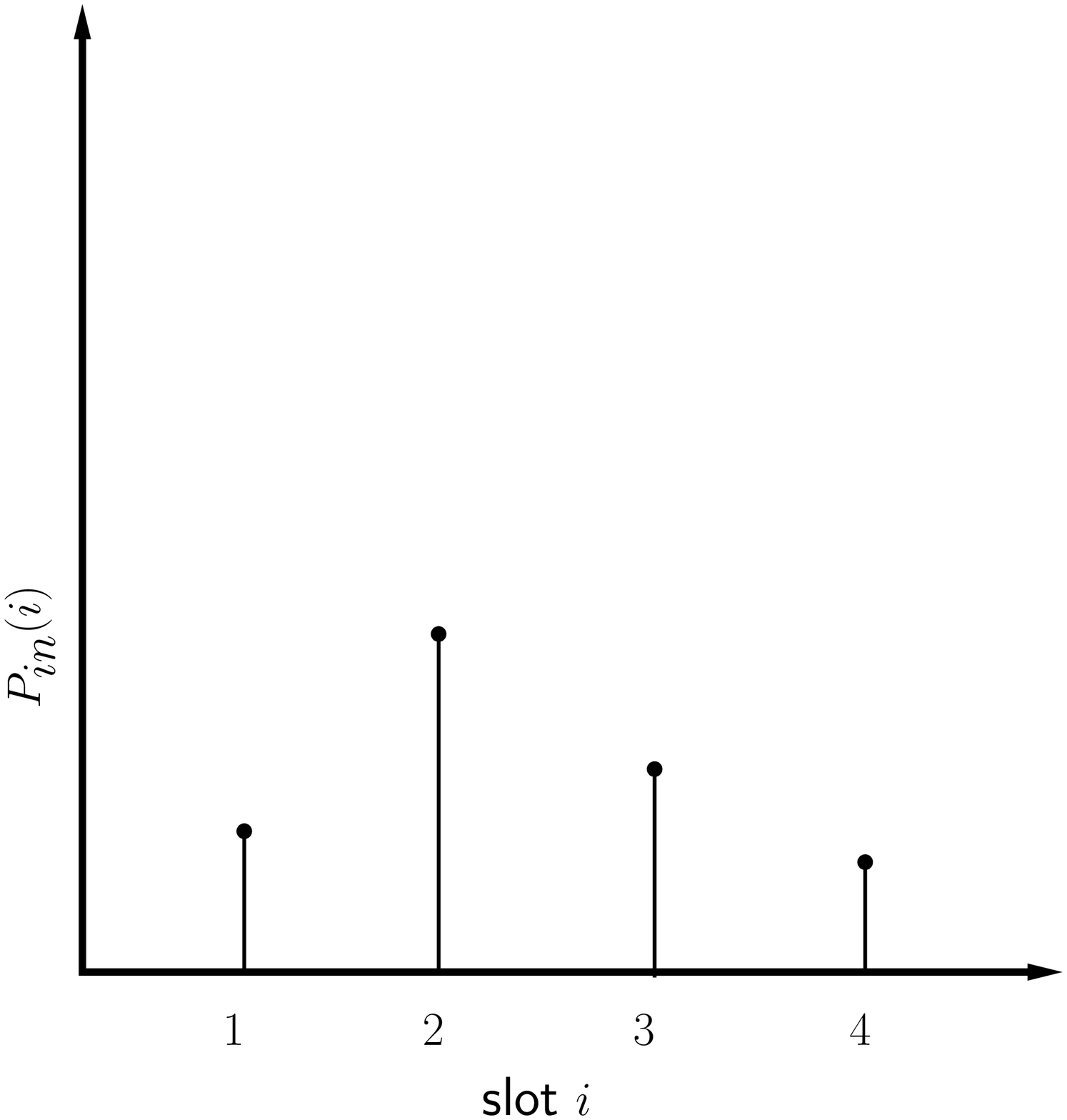}
\label{xx}
}
\subfigure{
\includegraphics[width=0.46\linewidth]{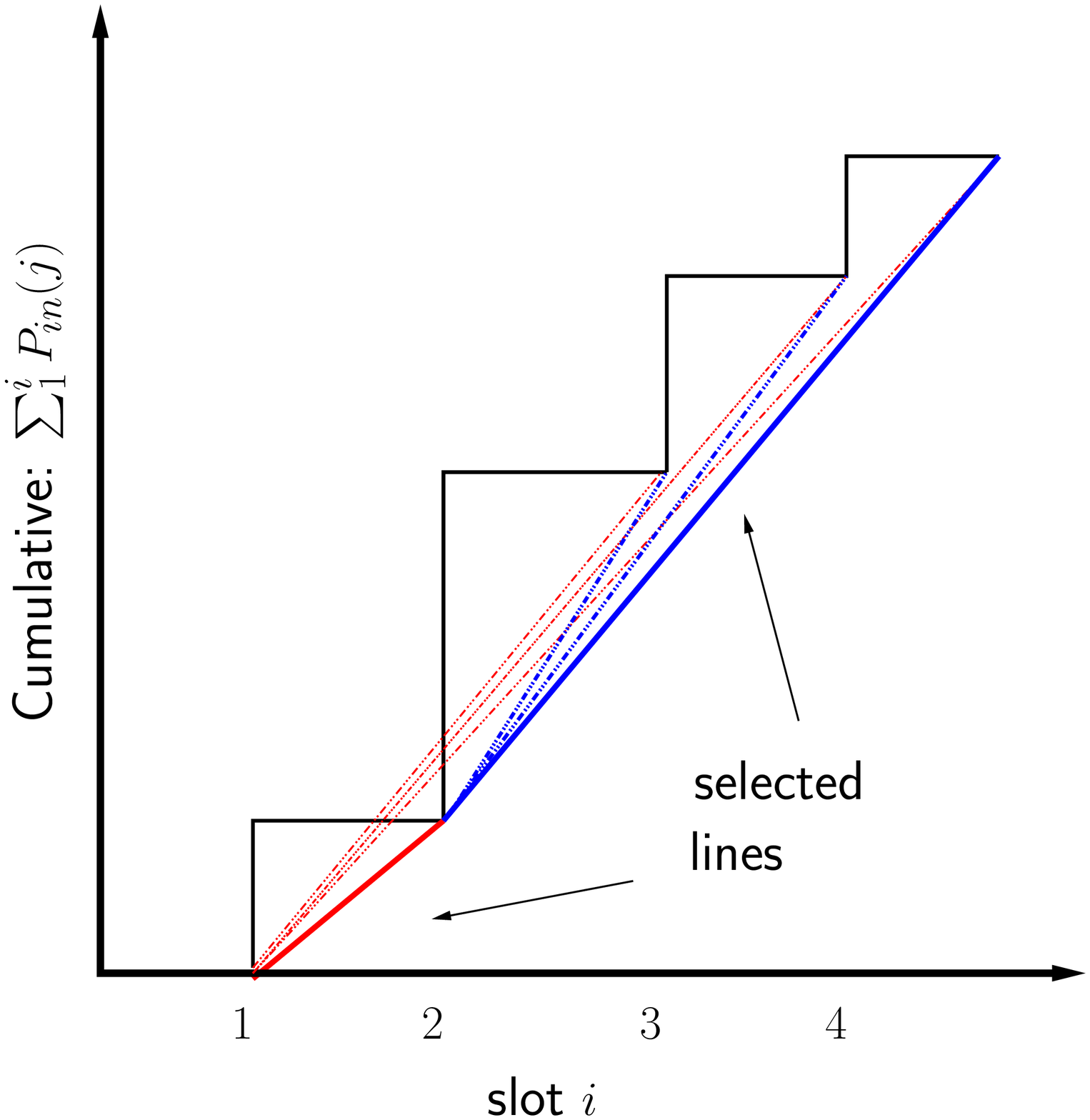}
\label{xy}
}
\caption{Operation of the optimal power allocation algorithm. On the left figure,
the power impulses are shown and on the right figure, cumulative energy arrival rate
is shown and construction of optimal power allocation as minimum feasible piecewise linear curve that is under the cumulative arrivals is illustrated. $P_{tr}^*(i)$ is the slope of the selected line in slot $i$.}
\label{opalg}
\end{figure}

An illustration of the operation of the algorithm is presented
in Fig. \ref{opalg}. The operation of the algorithm has the
following geometric structure. At each point $i$, lines are
drawn from cumulative energy point $(i,e(i))$ to future points
$(k,e(k))$ for all $k>i$ and the one with minimum slope is
chosen, say $k^*$. That slope is the allocated power for all slots between $i$ and $k^*$.

\subsection{A Numerical Study of the Algorithm}

The optimal power management algorithm takes the arrival power
vector $[P_{in}(1),...,P_{in}(L)]$ and outputs the transmit power vector $[P_{tr}^*(1),...,P_{tr}^*(L)]$. We let the arrival rates of
energy in all slots, $P_{in}(i)$, follow an i.i.d. exponential
distribution.

A benchmark algorithm is simply no power management algorithm,
i.e., $P_{tr}(i)=P_{in}(i)$. In this simple scheme, the energy
arrival rate in each slot is taken as the communication power
in that slot. This scheme yields an average throughput
\begin{equation}
T^{lb}= \frac{1}{L} \sum_{i=1}^{L}\frac{1}{2}\log\left(1 +
P_{in}(i)\right)
\end{equation}
which is a lower bound. However, if the designer has the
information of arrival rates in future slots, then optimal
power management algorithm can improve the average throughput.
It is clear that an upper bound for the average throughput is
\begin{equation}
T^{ub}= \frac{1}{2}\log\left(1 + \frac{1}{L} \sum_{i=1}^{L}
P_{in}(i)\right)
\end{equation}

\begin{figure}
\begin{center}
\includegraphics[height=0.6\linewidth]{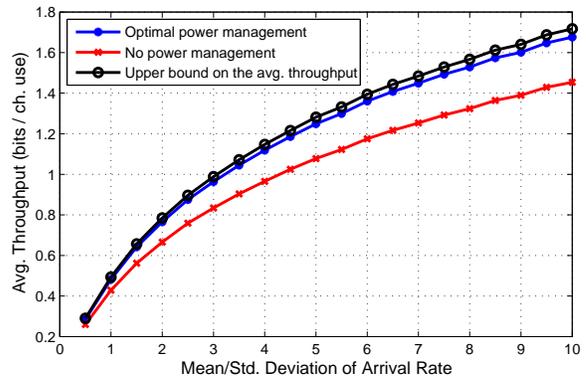}
\end{center}
\caption{Average throughput in $L=20$ slots is plotted for different
values of mean/standard deviation of the arrival rate. As the variation
in the arrival rate increases, optimal power management algorithm yields
higher advantage over no power management.}
\label{numres}
\end{figure}

The comparison of performances of optimal power management with
the upper bound $T^{ub}$ and the lower bound $T^{lb}$ (no power
management) is given in Fig.~\ref{numres} for a $L=20$ slot
system. We observe that as the variance of the arrival rates
increases, the advantage of optimal power management becomes
more apparent with respect to no power management. Because, the
power is more peaky as the variance is increased yet for better
throughput, transmitter should not ride the peaks but rather
save the energy for future use. Another observation is that the
difference between upper bound and average throughput with
optimal power management also increases as the standard
deviation of the arrival rate is increased. Hence, the
causality constraint becomes more restrictive as the variation
in the arrival rate is increased.

%

\section{Conclusion} \label{cnc}

We studied communication in an AWGN channel under random energy arrivals using an
information-theoretic framework. We showed that the capacity of AWGN under energy
harvesting is the same as the capacity of the AWGN channel, where the average power is constrained to the
average recharge rate. This upper bound can be achieved by save-and-transmit
and best-effort-transmit schemes. Next, we addressed time varying recharge rates in large time scales. We
obtained an algorithm to find the optimal power management for maximum average
throughput. We illustrated its operation and provided a numerical study to visualize the advantage of the algorithm.

\bibliographystyle{ieeetr}

\end{document}